# Observation of non-planar dust acoustic solitary wave in a strongly coupled dusty plasma

Prarthana Gogoi [a), b)], Bidyut Chutia [c)], Paragjyoti Sut [a), b)], Yoshiko Bailung [d)], Nirab C. Adhikary [a)] and H. Bailung [a), b), e)]

a) Physical Sciences Division, Institute of Advanced Study in Science & Technology (IASST), Paschim Boragaon, Guwahati 781035, Assam, India
b) Academy of Scientific and Innovative Research (AcSIR), Ghaziabad – 201002, India
c) Centre of Plasma Physics – Institute for Plasma Research (CPP-IPR), Nazirakhat, 782402, Assam, India
d) Department of Physics, Goalpara College, Goalpara, 783101, Assam, India
e) Department of Physics, Bodoland University, Kokrajhar 783370, Assam, India (on lien from IASST)

**Abstract:**

The nonlinear evolution and propagation of a stable dust acoustic solitary wave (DASW) in a non-planar geometry is investigated here. The experiment is performed in a strongly coupled dusty plasma consisting of monodisperse micron particles levitated in the sheath of a capacitively coupled RF argon plasma. The non-planar waves are generated with the help of a cylindrical conducting exciter pin placed at the centre of the homogeneous dust cloud. A negative excitation pulse is used to create a dust void and a dust density perturbation simultaneously around the exciter. From the edge of the void, the density perturbation propagates as a nonlinear (cylindrical) non-planar dust acoustic solitary wave. The characteristics of the solitary wave are measured using image analysis of the recorded video of wave propagation. Numerical solution of the modified Korteweg de Vries (KdV) equation with an additional term to take care of the non-planar geometry is compared with the experimental observation. The wave amplitude and width are measured as a function of time and compared with the theoretical predictions.

**Introduction:**

A dusty plasma is a unique plasma medium that consists of macroscopic dust particles (highly charged) embedded in a conventional plasma environment of electrons, ions, and neutrals. Dusty plasmas exist throughout our universe, like the interstellar cloud, interior of stars, nebulas, planetary rings, and cometary tails etc.[1–4]. In laboratories they appear in industrial plasma processing reactors[5] and in thermonuclear fusion devices like tokamaks[6]. Inside the plasma, the micron size dust particles acquire a huge amount of charge due to inflow of electrons and ions into its surface. However, owing to the higher thermal velocity of electrons dust particle mostly get negatively charged, resulting in the modification of charge neutrality condition. These charged dust particles interact with each other via Coulomb force (and get screened by the background plasma electrons and ions) and the strength of this interaction determines the state of the dusty plasma medium. A dusty plasma is in the strongly (weakly) coupled state when the ratio of the electrostatic Coulomb potential energy to the average thermal energy of the dust particles is greater (less) than 1. When this ratio exceeds a certain critical value (~170), the dust particles arrange themselves in ordered structure like crystals[7,8].

In the past couple of decades, studies on dusty plasma in laboratory gained prominence on account of their ability to support a wide variety of collective phenomenon associated with linear and non-linear waves. The presence of these massive charged dust particles in the medium gave rise to a new wave mode known as dust acoustic wave (DAW) [9] in addition to modification of the existing plasma wave modes (such as ion acoustic wave gets modified into dust-ion acoustic wave)[10,11]. The DAW is a low frequency analog of the ion acoustic wave mode, wherein the charged dust particles provide the inertia, and the electrons and ions provide the restoring force. Initially, the low frequency dust acoustic waves have been found to excite spontaneously in variety of dusty plasma experiments which are in conformity with the theoretical predictions[12,13]. Since then, a large number of investigations have been carried out to study the evolution and propagation of the DAW in both earth-based laboratories[14–17] as well as under microgravity conditions[18–20].

The nonlinear evolution of a DAW into unidirectional soliton, propagating with near-sonic velocities was also presented by Rao in Ref. 9. Solitons are special class of non-linear waves that can preserve their shapes and velocities over a long distance while traversing through a medium on account of their ability to strike a delicate balance between the nonlinear wave steepening and dispersive wave broadening. Another unique property of a solitary wave is its ability to preserve its identity on collision with another solitary wave on account of which they are termed as solitons [21]. These long-standing waves which were first observed in water by J Scott Russell [22], has since been studied in variety of physical media, including the Jovian magnetosphere[23], Earth's magnetopause [24], ocean waves [25], propagation of action potential along a neuron's axon [26,27], signal transmission in optical fibres [28] as well as a host of other fluid mediums.

Samsonov *et al*. reported the first experimental observation of solitons in a 2D monolayer hexagonal dust lattice and explained their findings using the linear chain theory[29]. DASW has also been excited in a 2D strongly coupled dusty plasma using laser excitation technique and the characteristics of the solitons have been found to follow the relationship between amplitude, width and velocity described by the KdV equation [30]. An experiment on head on collision between two counter propagating DASW in a strongly coupled dusty plasma revealed that they merge into a single pulse during collision and then pass through each other with a slight time delay [31]. In another study, wave breaking of a self-excited dust acoustic soliton has been observed and found that the breaking of a soliton can accelerate the particles to supersonic speeds [32]. In a recent experiment, reflection of a DASW from a potential barrier has been reported[33]. However, all of the study conducted on DASW till date has been restricted to one dimensional propagation. Theoretical studies focusing on how the non-planar geometry effects the propagation of ion acoustic solitary waves have been carried out within the realm electron-ion plasma using a modified KdV equation[34,35]. This modified KdV equation contains an additional term which describes the geometric divergence or convergence of spherical and cylindrical wave fronts. On account of the presence of this additional geometrical term, a stationary asymptotic solution of the equation does not exist. However, an approximate analytical expression has been obtained for early development of ingoing solitons along with their numerical solutions, which is found to differ from their one-dimensional counterpart. Experimentally, ion acoustic solitons are observed in non-planar geometry (spherical) in a collisionless plasmas[36].

Mamun and Shukla [37] presented a theoretical investigation on the cylindrical and spherical dust acoustic solitary wave and derived the modified KdV equation with an additional geometrical term. The medium they considered was unmagnetized dusty plasma with cold negatively charged dust fluid and Boltzmannian electrons and ions. Their approach was similar to the one used by Maxon and Viecelli[35] for cylindrical ion acoustic solitons. Numerical solutions for the same are obtained using a two-level finite difference approximation method and the propagation characteristic of the higher dimensional waves are found to differ significantly from their one-dimensional counterpart.

Here we report the experimental observation of a non-planar dust acoustic solitary wave in a strongly coupled dusty plasma fluid. The manuscript is organized as follows. The theory and numerical results based on the non-planar KdV equation is presented in section II. In section III we discuss the experimental set-up and procedure. The results and discussion are presented in section IV. Section V contains the conclusion.

**Theory and Numerical Results:**

We consider a modified KdV equation with an additional geometrical term derived for an unmagnetized dusty plasma whose constituents are negatively charged cold dust fluid and Boltzmann electrons and ions and is given as [37]

$$\frac{\partial n_d}{\partial \tau} + \frac{v}{2\tau} n_d - A n_d \frac{\partial n_d}{\partial \zeta} + B \frac{\partial^3 n_d}{\partial \zeta^3} = 0 \qquad (1)$$

Where $v = 0, 1, 2$ for planar, cylindrical, and spherical geometries, respectively. $n_d$ is the dust density perturbation normalized by the equilibrium dust density $n_{d0}$. Equation (1) is similar to the KdV equation except for the second term that accounts for the geometric divergence or convergence of the wave. The parameters $\tau$ and $\zeta$ are the stretched time and space coordinates respectively given as $\tau = \varepsilon^{3/2} t$ and $\zeta = -\varepsilon^{1/2}(r + u_0 t)$, where $\varepsilon$ is the smallness parameter and $u_0$ is the wave phase velocity normalized by the dust acoustic velocity $C_{DA} = \omega_{pd} \lambda_{Dd}$. The time and space variables are normalized by the inverse of the dust plasma frequency, $\omega_{pd}^{-1} = (m_d / 4\pi n_{d0} Z_d^2 e^2)^{1/2}$ and dust Debye length $\lambda_{Dd} = (\lambda_{De}^{-2} + \lambda_{Di}^{-2})^{-1/2}$, respectively, where $\lambda_{De} = (T_e / 4\pi n_{e0} e^2)^{1/2}$ is the electron Debye length and $\lambda_{Di} = (T_i / 4\pi n_{i0} e^2)^{1/2}$ is the ion Debye length, $m_d$ is the dust mass, $n_{e(i)0}$ is the equilibrium electron (ion) density, $Z_d$ is the average charge on a dust particle, and $T_{e(i)}$ is the electron (ion) temperature. The coefficients for nonlinearity (A) and dispersion (B) in Eq. (1) are expressed as $A = \frac{u_0^3}{(\delta - 1)^2} \left[ \delta^2 + (3\delta + \sigma_i)\sigma_i + \frac{1}{2}\delta(1 + \sigma_i^2) \right]$ and $B = \frac{u_0^3}{2}$ respectively. Here $\delta$ and $\sigma_i$ are the ratios of ion to electron number density and temperature respectively, i.e., $\delta = n_{i0}/n_{e0}$ and $\sigma_i = T_i/T_e$.

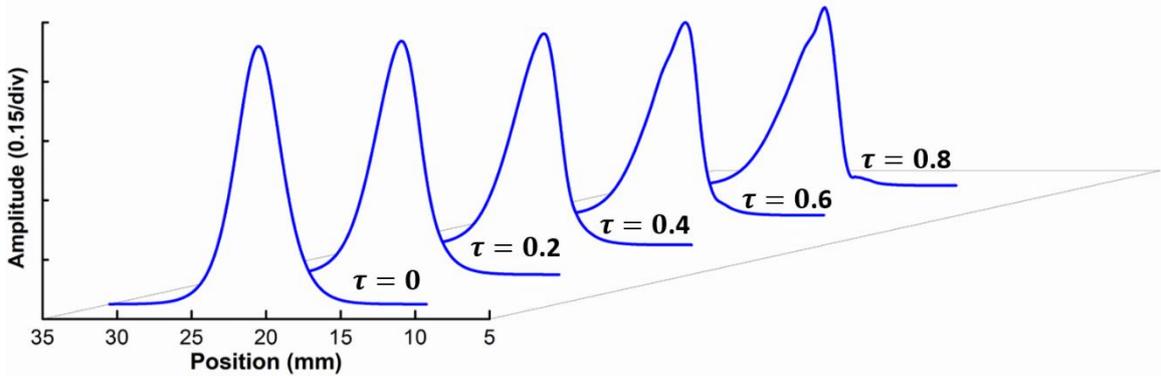

FIG. 1: Solitary wave profile obtained from the numerical solution of the modified KdV equation with an addition geometrical term $(v/2\tau)n_d$, for $v = 1$.

The modified KdV equation (1) is solved numerically for cylindrical geometry ($v = 1$). A stationary solution of equation (1) $n_d(\zeta, \tau) = n_{dm} sech^2(\zeta/\Delta)$ as the initial condition (with $v = 0$ for the second term) is used to observe the evolution. Here $n_{dm} = 3/A$ is the amplitude of the solitary wave and $\Delta = \sqrt{4B}$ is its width. Fig. 1 shows the space evolution of the solitary wave as it propagates through the medium at different times. The values of the parameters used in the theoretical calculations are $u_0 = 0.3$, $\sigma_i = 0.02$, and $\delta = 1.1$ from which the values of the coefficients A (nonlinear) and B (dispersion) are estimated. Due to higher initial amplitude the perturbations beyond $\tau = 0.4$, show the appearance of the second solitary wave behind (on the shoulder) the first peak.

**Experimental Set-up:**

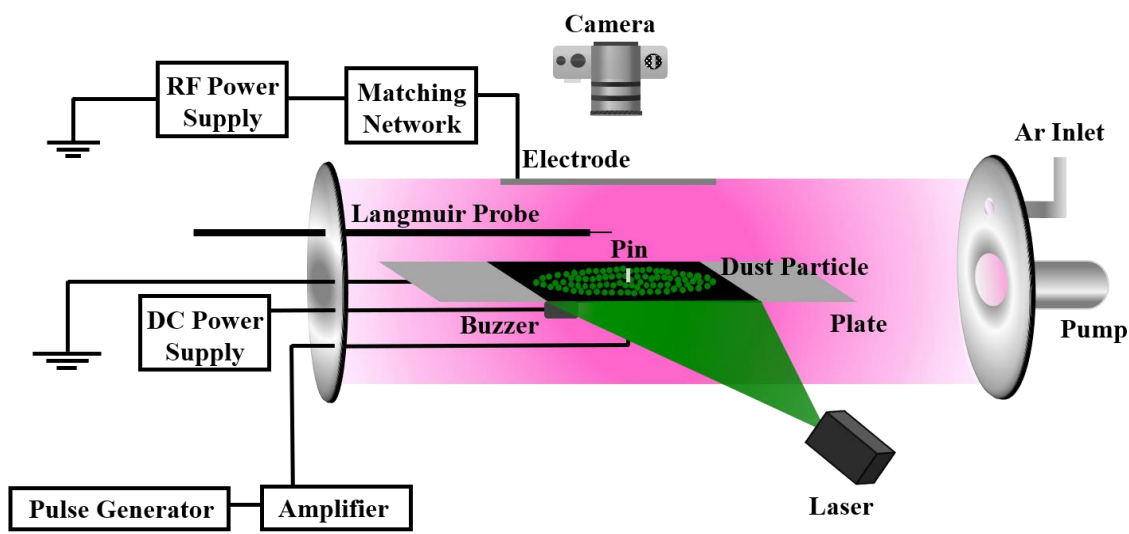

FIG. 2: Schematic diagram of the experimental set-up.

The experiment has been carried out in a cylindrical glass chamber (DuPLEX) of length 100cm and meter 15 cm [38,39]. A schematic diagram of the experimental setup is shown in Fig. 2. Initially, the chamber is evacuated down to a base pressure of the order of ~$10^{-5}$ mbar and then argon gas is slowly injected using a precision needle valve to reach working pressure of ~$10^{-2}$ mbar. The plasma is then produced by applying radio frequency (RF) power (13.56 MHz, 5W) via a matching network to two thin parallel aluminium strips (each of length 20 cm and width 2.5 cm) placed on the outer surface of the chamber. A rectangular stainless-steel (SS) plate of length 30 cm and breadth 14.5 cm is placed horizontally inside the chamber which is kept grounded. On the plate a small hole of diameter ~0.3 cm is made below which a piezo-electric buzzer filled with the dust particles is attached. The dust particles used are monodisperse gold coated silica particles of size ~5 μm and mass density ~2.6 gm cm$^{-3}$. After the production of the argon plasma, the dust particles are dispersed into the plasma by switching the piezo electric buzzer on. Inside the plasma, these particles get charged, mostly negatively, on account of the inflow of plasma electrons and ions. These massive (compared to electrons and ions) and highly charged (negatively) particles then levitate in the plasma sheath boundary region at about ~0.7 cm above the grounded SS plate at an equilibrium position where the downward gravitational force acting on them is balanced by the upward sheath electric field force. Two graphite strips are placed vertically on both ends of the plate to provide an electrostatic barrier to prevent particle loss. These charged microparticles are illuminated using a horizontal sheet of green laser light (50 mW, 532 nm Optoelectronics Technology Co. Ltd., China) and their dynamics is recorded using a high-resolution, high-speed camera (Phantom MIRO M110, USA) at 30 and 100 frames per second (fps).

In order to characterize the background Ar plasma, an RF compensated cylindrical Langmuir probe made up of tungsten wire of length ~0.5 cm and diameter ~0.02 cm connected to an Automated Langmuir Probe system (Impedans ALP 150, Ireland) is used. The typical values of the measured plasma parameters are as follows: ion density ~ $10^8$ cm$^{-3}$, electron temperature $T_e$ is ~5 eV, plasma potential ~ +35 V. The average charge $Q_d$ on a dust particle is determined from the force balance equation between the upward sheath electrostatic force $Q_d E_h$ (where $E_h$ is the electric field at dust levitation height $h$) and the downward gravitational force $m_d g$ (where $m_d = 1.7 \times 10^{-13}$ kg is the dust mass and $g$ is the acceleration due to gravity). Details of the measurement of the sheath electric field at dust levitation height has been described in Ref. 38. At typical dust levitation height of (0.7 – 1.0 cm) the average charge on a dust particle $Q_d (= m_d g / E_h)$ is measured to be ~$10^4$ e. The $Q_d$ value derived in this manner closely agrees with the dust charge value computed using the Orbital Motion Limited (OML) theory. The dust density $n_d$ which is estimated from the interparticle distance $a$ (using $n_d = 3/4\pi a^3$) is measured to be ~$10^3$ cm$^{-3}$. The Coulomb coupling parameter which describes the electrostatic interaction between neighbouring dust particles in terms of their thermal energy is calculated using the relation $\Gamma = (Z_d^2 e^2 / 4\pi\varepsilon_0 a \kappa_B T_d) exp(-a/\lambda_D)$, where $Z_d = (Q_d/e)$ is the dust charge number, $e$ is the electronic charge, $\varepsilon_0$ is the absolute permittivity, $a$ is the interparticle distance, $\kappa_B$ is the Boltzmann constant, $T_d$ is the dust temperature and $\lambda_{Dd}$ is the linearized dust Debye length. For the present experimental conditions, $\Gamma$ is found to be ~103 which is sufficiently larger than 1, however, smaller than the critical coupling constant $\Gamma_c$ ~170. Thus, the present experimental dusty plasma medium is in the strongly coupled regime.

**Non-planar wave excitation:**

In order to generate nonplanar density perturbation, a thin and sharp SS pin is used as an exciter. The pin is placed vertically at the mid portion of the grounded plate with proper insulation. The height of the pin is ~0.6 cm so that it lies just below the dust floating plane. An arbitrary function generator is used to apply pulsed DC signal to the exciter. In absence of any external bias, the exciter pin creates a circular dust void (~1 cm in diameter) with a sharp void boundary. The void boundary lies at the region where the outward electrostatic force exerted by the floating pin (negative with respect to the plasma potential) on dust particles is balanced by the inward ion drag force. A thorough investigation on the sheath structure and the force balance on dust particles under similar experimental conditions has been described in Ref. 41. The void appears because the exciter pin acquires a floating potential ~ +2 to 3 V, which is negative relative to the plasma potential (~ +35 V) A dc offset of ~ +32 V is applied to the exciter to bring the exciter (as well as the void region) to the same potential level by neutralizing the sheath effect. Thus, in absence of the sheath electric field, the dust particles fill the void region and a uniform dust cloud is formed around the exciter. Before the application of any external perturbation, the dusty plasma medium is maintained in stable equilibrium so that no spontaneous perturbations are excited. In order to excite the dust acoustic perturbation, a negative pulse of duration ~1 s and amplitude (~ few tens of volts) is applied to the exciter pin. As a result, the dust particles surrounding the pin moves outward forming a density compression and the void keeps on expanding. The expansion of the void ends at the sheath edge of the pin. From the edge of this stable void boundary, a diverging density perturbation propagates as a non-planar DASW. This entire process of the generation of the compressive density perturbation and evolution into a solitary wave transpires within ~140 ms after the application of the excitation pulse.

**Experimental Results and Discussion:**

Typical raw data (snapshots) representing the excitation and propagation of the nonplanar (cylindrical) dust density perturbation is shown in fig. 3. The snapshots are extracted from video recorded at 100 frames per second with inter-frame timing of 10 ms. The excitation pulse ( -60 V, 1.2 s duration) is applied to the exciter pin (located at the centre of the void) at

t =0. Three snapshots in the top row show the evolution of the circular dust void simultaneously with the dust density compression at three different times (0.01 – 0.10 s).

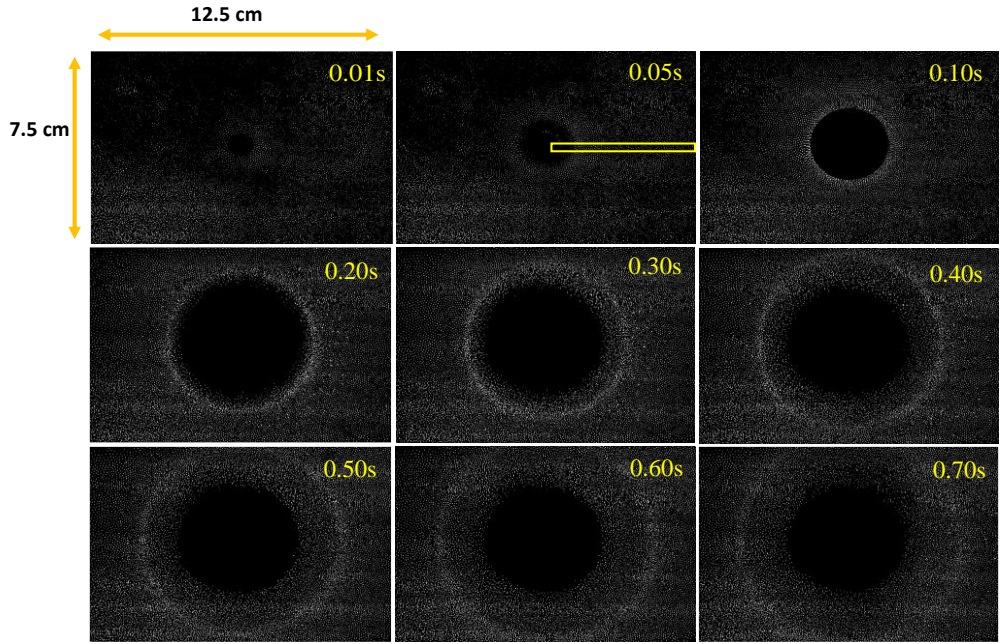

FIG. 3: Snapshots depicting the evolution of dust void and the diverging dust density perturbation. The exciter pin is at the centre of the void. The thin rectangular box shown in the image ($t = 0.05\ s$) represents a typical region considered for analysis of image frames. Horizontal stripes in the lower part of image frames are appearing due to reflection of laser from the boundary of the glass chamber.

Initial examination on the temporal evolution of the dust void (size) is found to be similar in nature as theoretically reported[40]. It is interesting to note that dust void expansion rate in the final phase coincides with the velocity of the density perturbation which starts to propagate in the dusty plasma medium from the void boundary. A detail investigation on such externally generated void evolution and its role in excitation of nonplanar density perturbation will be taken up in later stage.

The void expansion ceases at ~140 ms after reaching maximum diameter of ~4.0 cm. Fig. 4 shows the nonlinear nature of the growth of the void size with time in response to the sudden application of the negative dc pulse at the exciter pin. The void expansion stops at the circular boundary where the force balance occurs between sheath electric field force (outward) and counteracting ion drag force into the sheath (towards the void). The dependence of void size with the bias applied to the pin has been investigated earlier[41]. From this void boundary the dust density perturbation propagates through the dust medium which is depicted in snapshots (second and third row) at constant time interval of 100 ms. Snapshot at t = 0.2 s shows initial separation of the density perturbation from the void boundary.

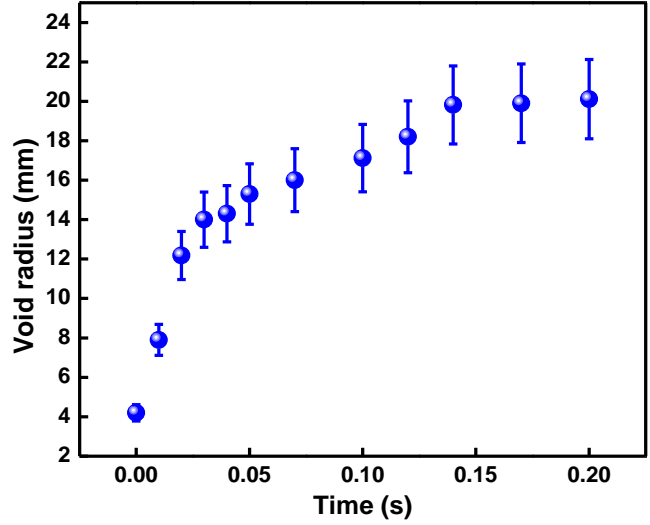

FIG. 4: Radius of the dust void as a function of time.

With increasing time, the density perturbation propagates through the dusty plasma medium keeping its cylindrical shape symmetrically outwards from the void. The amplitude and the width of the perturbation are observed to be modified with time. The excitation voltage is far above the threshold value and hence the density perturbation shown in the dataset is in the nonlinear regime. The dust density perturbation propagates keeping its nonplanar cylindrical shape expanding through the dusty plasma medium which is followed up to ~4 cm from the void boundary. A typical 3-D pixel intensity profile of the cylindrically propagating perturbation corresponding to time 300 ms is shown in fig. 5. The blue cylindrical front (in z-axis) represents the compressive perturbation on the green background of unperturbed dusty plasma medium. The red dip circle in the mid portion represents the dust void.

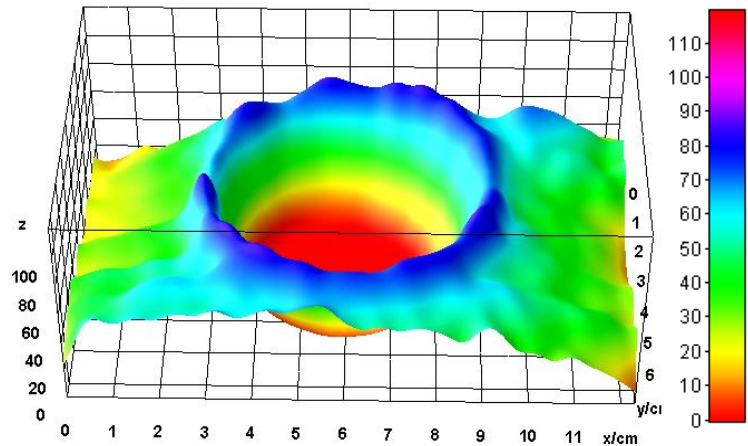

FIG.5: A typical 3-D pixel intensity profile of the cylindrical DASW created from experimental image frame using ImageJ software. The time is 0.3s after application of excitation pulse. The colour bar represents the amplitude of the perturbation. The x and y axis are representative of the length and breadth of the raw image whereas the z axis represents its pixel intensity.

In order to understand evolution of the compressional longitudinal density perturbation over time, the pixel intensity profile of the snapshots is measured along a radius considering a small rectangular section. The pixel intensity is calibrated with the normalized dust density perturbation $[(n_d/n_{d0}) - 1]$. A typical set of pixel intensity profiles corresponding to the dataset (shown in Fig.3) is presented in Fig. 6. The location of the exciter pin at the centre of the void correspond to zero on the position axis. In order for the compressional density perturbation to evolve into a solitary wave structure, the excitation pulse applied is maintained above certain threshold value. This is because a unipolar, longitudinal pulse while propagating through the medium containing charged dust particles will undergo dispersive broadening. Hence to counteract it, the amplitude of the excitation pulse must be high enough so that the perturbation can propagate to a sufficient distance before the non-linear steepening gets balanced by the dispersive effects of the medium. This delicate balancing between the two-contrasting effects ultimately lead to the formation of a stable solitary wave. In the present experiment, the evolution of the dust density perturbation into a solitary wave occurs mostly inside the void region (diameter ~4.0 cm) where both the perturbation and the dust medium move along simultaneously.

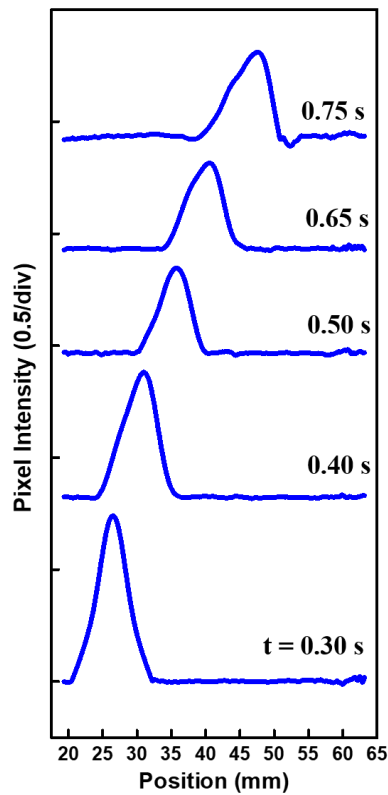

FIG.6: Pixel intensity profiles depicting the spatial evolution of cylindrical dust acoustic solitary wave at different times

Interestingly the perturbation is found to take solitary wave of secant hyperbolic squared form when it enters the dusty plasma medium after the void region. As time goes by the amplitude of the perturbation decreases and width increases. In later time at far away locations indication of formation of the second solitary wave is seen on the shoulder of the first peak (t = 0.65 s and 0.75 s).

The experimental results are therefore observed to have a close resemblance with the structures of wave profiles obtained from the numerical solution of the KdV equation modified for cylindrical geometry (Fig. 1).

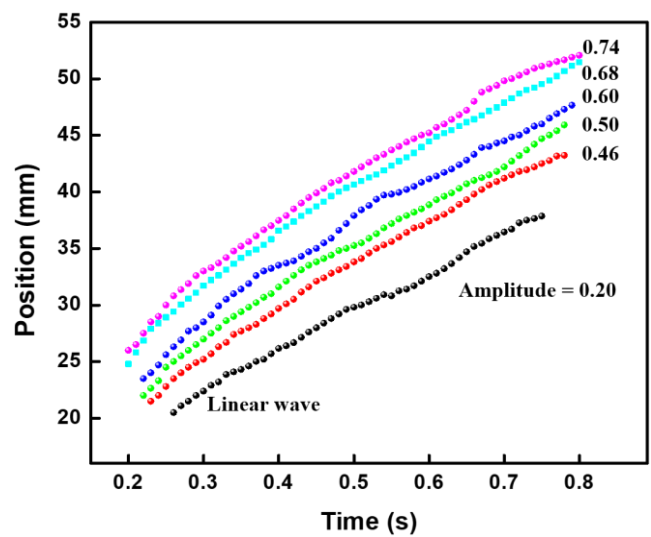

FIG.7: Trajectories incurred by the linear DAW and the cylindrical DASWs for different normalized perturbation amplitude whose values are depicted by the numbers presented to the right of the trajectories.

The space-time trajectories of the perturbation propagating with their cylindrical shape through the medium are shown in Fig.7 along with its linear counterpart (at the bottom). The linear perturbation is corresponding to a very small amplitude ~ 0.20 (normalised amplitude). Extreme care is taken to follow the perturbation with lowest detectable amplitude. The measured velocity of the linear perturbation is ~3.4 cm s$^{-1}$. The slopes of the space time trajectories of the perturbations increase with increasing amplitude indicating the increase in the wave velocity. The measured velocities of the perturbation are found to lie between ~3.9 cm s$^{-1}$ to ~4.2 cm s$^{-1}$ when normalized amplitude is varied from 0.46 to 0.74.

In Fig. 8, the Mach numbers of the cylindrical dust acoustic perturbations are plotted as a function of normalized wave amplitude. The Mach velocity increases with increase in the wave amplitude. The measured Mach velocities range from 1.15 to 1.24 when the amplitude is increases from 0.46 to 0.74.

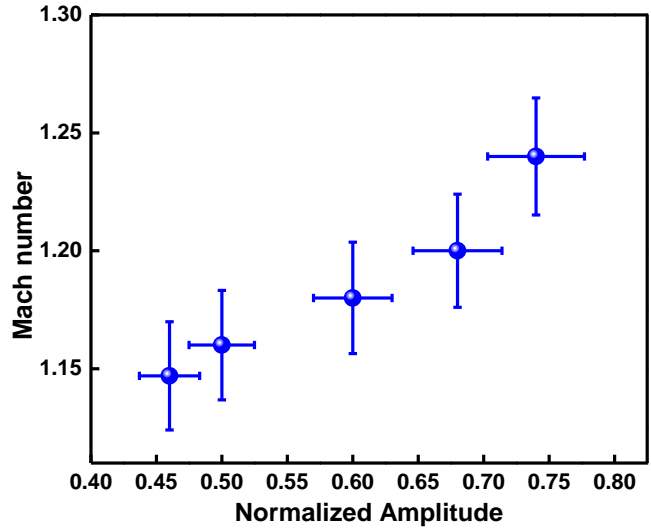

FIG 8: Mach number variation with amplitude of cylindrical DASW.

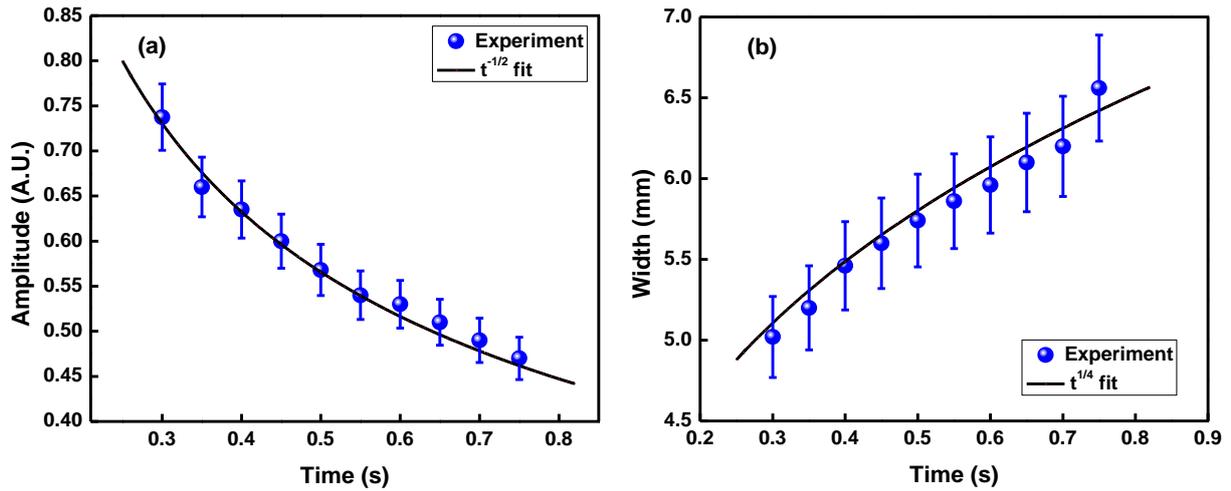

FIG 9: Variation of the (a) amplitude and (b)width of the solitary wave as a function time.

The variations of amplitude and width of the cylindrical DASW with time are shown in Fig. 9 (a) and (b) respectively. The width is calculated by measuring the full width at half maximum of the perturbation represented by the pixel intensity. Fig. 9(a) indicates decrease in the wave amplitude from 0.74 (at time, t =0.30 s) to 0.47 (at time, t =0.75 s), as it propagates through the medium. Concurrently, within the same time interval, the width of the perturbation increases from 5.1 mm to 6.6 mm, which is shown in Fig. 9(b). This nonlinear variation of the experimentally measured amplitude and width of the wave are empirically scaled as being proportional to $t^{-1/2}$ and $t^{1/4}$ respectively. These are special characteristics of cylindrical ion acoustic solitary waves obtained theoretically by Maxon and Viecelli [35] which are also followed by the nonlinear dust acoustic perturbations observed in the present experiment.

**Conclusion:**

To summarize, we have carried out an experimental investigation on the evolution and propagation of a non-planar DASW in a homogeneous, strongly coupled dusty plasma in a capacitively coupled RF argon plasma. The non-planar waves are generated with the help of a specially designed exciter (a cylindrical pin of very small diameter) placed at the centre of the dust cloud to which a negative dc pulse is applied. On the application of the excitation pulse, the sheath around the pin expands creating a dust void around itself and a diverging cylindrical dust density perturbation. The perturbation develops as a longitudinal density compression within the sheath region (later become void) and emerges as a single diverging dust acoustic solitary wave propagating through the dusty plasma medium. The density perturbations are followed with high-speed camera using laser illumination. The amplitude, width and the Mach number of the perturbations are measured as a function of time. The nonplanar dust acoustic solitary waves are found to follow a similar characteristic (empirical relationship) as predicted from theoretical analysis for nonplanar ion acoustic solitary waves in Ref. 35. Experimental wave profiles are found to closely resemble with the structure obtained from the numerical solutions of the modified KdV equations with an addition geometrical term. Our study conducted under controlled laboratory conditions, could provide one with valuable insights into understanding the fundamental characteristics of a non-planar solitary wave and should be useful in future investigations of such phenomena under laboratory as well as space plasma environments. Furthermore, the outcomes of our investigations can provide a template for comprehending these waves in other media, where the ability to control the different parameters might be limited and understand them at the kinetic level using relatively simpler methods.

**Author Declarations:**

**Conflict of Interest**

The authors have no conflict of interest to disclose.

**References**

[1] C. K. Goertz, Rev. Geophys. **27**, 271 (1989).

[2] M. Tátrallyay, M. Horányi, A. Juhász, and J. G. Luhmann, Adv. Sp. Res. **12**, 27 (1992).

[3] T. W. Hartquist, W. Pilipp, and O. Havnes, Astrophys. Space Sci. **246**, 243 (1997).

[4] C. J. Mitchell, M. Horanyi, O. Havnes, and C. C. Porco, Science **311**, 1587 (2006).

[5] L. Boufendi, and A. Bouchoule, Plasma Sources Sci. Technol. **11**, A211 (2002).

[6] J. Winter, Plasma Phys. Control. Fusion **40**, 1201 (1998).

[7] H. Ikezi, Phys. Fluids **29**, 1764 (1986).

[8] H. Thomas, G. E. Morfill, V. Demmel, J. Goree, B. Feuerbacher, and D. Möhlmann, Phys. Rev. Lett. **73**, 652 (1994).

[9] N. N. Rao, P. K. Shukla, and M. Y. Yu, Planet. Space Sci. **38**, 543 (1990).

[10] P. K. Shukla, and V. P. Silin, Phys. Scr. **45**, 508 (1992).

[11] A. Barkan, N. D'Angelo, and R. L. Merlino, Planet. Space Sci. **44**, 239 (1995).

[12] N. D'Angelo, J. Phys. D. Appl. Phys. **28**, 1009 (1995).

[13] A. Barkan, R. L. Merlino, and N. D'Angelo, Phys. Plasmas **2**, 3563 (1995).

[14] C. Thompson, A. Barkan, N. D'Angelo, and R. L. Merlino, Phys. Plasmas **4**, 2331 (1997).

[15] T. Trottenberg, D. Block, and A. Piel, Phys. Plasmas **13**, 042105 (2006).

[16] E. Thomas, Jr., Phys. Plasmas **13**, 042107 (2006).

[17] B. Chutia, T. Deka, Y. Bailung, D. Sharma, S. K. Sharma, and H. Bailung, Phys. Plasmas **28**, 123702 (2021).

[18] A. Piel, M. Klindworth, O. Arp, A. Melzer, and M. Wolter, Phys. Rev. Lett. **97**, 205009 (2006).

[19] M. Schwabe, S. K. Zhdanov, H. M. Thomas, A. V. Ivlev, M. Rubin-Zuzic, G. E. Morfill, V. I. Molotkov, A. M. Lipaev, V. E. Fortov, and T. Reiter, New J. Phys. **10**, 033037 (2008).

[20] C. A. Knapek, M. Schwabe, V. Yaroshenko, P. Huber, D. P. Mohr, and U. Konopka, Phys. Plasmas **30**, 033703 (2023).

[21] N. J. Zabusky, and M. D. Kruskal, Phys. Rev. Lett. **15**, 240 (1965).

[22] J. S. Russell, Report on Waves, 14th Meeting of the British Association for the

Advancement of Science, York, 311–390 (1845).

[23] G. Slathia, K. Singh, and N. S. Saini, J. Astrophys. Astron. **43**, 1 (2022).

[24] R. Trines, R. Bingham, M. W. Dunlop, A. Vaivads, J. A. Davies, J. T. Mendonça, L. O. Silva, and P. K. Shukla, Phys. Rev. Lett. **99**, 205006 (2007).

[25] A. L. New, and R. D. Pingree, Deep Sea Res. Part A, Oceanogr. Res. Pap. **37**, 513 (1990).

[26] O. V. Aslanidi, and O. A. Mornev, JETP Lett. **65**, 579 (1997).

[27] R. R. Poznanski, L. A. Cacha, Y. M. S. Al-Wesabi, J. Ali, M. Bahadoran, P. P. Yupapin, and J. Yunus, Sci. Rep. **7**, 2746 (2017).

[28] H. A. Haus, and W. S. Wong, Rev. Mod. Phys. **68**, 423 (1996).

[29] D. Samsonov, A. V. Ivlev, R. A. Quinn, G. Morfill, and S. Zhdanov, Phys. Rev. Lett. **88**, 095004 (2002).

[30] T. E. Sheridan, V. Nosenko, and J. Goree, Phys. Plasmas **15**, 073703 (2008).

[31] S. K. Sharma, A. Boruah, and H. Bailung, Phys. Rev. E **89**, 013110 (2014).

[32] F. M. Trukhachev, M. M. Vasiliev, O. F. Petrov, and E. V. Vasilieva, Phys. Rev. E **100**, 063202 (2019).

[33] K. Kumar, P. Bandyopadhyay, S. Singh, G. Arora, and A. Sen, Phys. Plasmas **28**, 103701 (2021).

[34] S. Maxon, and J. Viecelli, Phys. Rev. Lett. **32**, 4 (1974).

[35] S. Maxon, and J. Viecelli, Phys. Fluids **17**, 1614 (1974).

[36] Y. Nakamura, M. Ooyama, and T. Ogino, Phys. Rev. Lett. **45**, 1565 (1980).

[37] A. A. Mamun, and P. K. Shukla, Phys. Lett. A **290**, 173 (2001).

[38] S. K. Sharma, R. Kalita, Y. Nakamura, and H. Bailung, Plasma Sources Sci. Technol. **21**, 045002 (2012).

[39] A. Boruah, S. K. Sharma, H. Bailung, and Y. Nakamura, Phys. Plasmas **22**, 093706 (2015).

[40] V. N. Tsytovich, S. V. Vladimirov, G. E. Morfill, and J. Goree, Phys. Rev. E **59**, 7055 (1999).

[41] Y. Bailung, T. Deka, A. Boruah, S. K. Sharma, A. R. Pal, J. Chutia, and H. Bailung, Phys. Plasmas **25**, 053705 (2018).